\documentclass[CEJP,PDF]{cej} 
\usepackage{graphicx}
\usepackage{layout}
\usepackage{amsmath}
\usepackage{textcomp}
\usepackage{hyperref}
\title{Comparison of Viscosities from the Chapman-Enskog and Relaxation Time Methods}

\articletype{Editorial}

\author{Anton Wiranata\inst{1}$^,$\inst{2}\email{aw125005@ohio.edu},
        Madappa Prakash\inst{3}\email{prakash@harsha.phy.ohiou.edu},
        Purnendu Chakraborty\inst{4}\email{purnendu.chakraborty@gmail.com}}

\institute{
     \inst{1} Central China Normal University,
     IOPP, Wuhan 430079, China
     \inst{2} Lawrence Berkeley National Laboratory,
     Nuclear Science Division, MS 70R0319, Berkeley, CA 94720, USA
     \inst{3} Ohio University,
     Department of Physics \& Astronomy, Athens, OH 45701, USA
     \inst{4} Physical Research Laboratory, Ahmedabad, India.
          }

\abstract{A quantitative comparison between the results of
shear viscosities from the Chapman-Enskog and relaxation time methods
is performed for selected test cases with specified elastic differential cross
sections:
(i) the non-relativistic, relativistic and ultra-relativistic hard sphere gas with
angle and energy independent
differntial cross section,
(ii) the Maxwell gas,
(iii) chiral pions and
(iv) massive pions.
Our quantitative results reveal that the extent of agreement (or
disagreement) depends very sensitively on the energy dependence of the
differential cross sections employed. }

\keywords{Shear Viscosity \*\ Chapman-Enskog Approximation \*\ Relaxation Time Approximation}
\pacs{01.30.-y, 01.30.Xx, 01.30.Tt}

\begin{document}
\maketitle

\section{Introduction}

The interpretation of the measured elliptic and higher order collective flows in terms of viscous hydrodynamics relies sensitively on the ratio of shear viscosity to entropy density.
Here, we quantify the extent to which results from
different approaches for shear viscosities of hadrons agree (or
disagree) by choosing some classic examples in which the elastic
scattering cross sections are specified. The two different
approximation schemes chosen for this study are the Chapman-Enskog and
relaxation time methods. The test cases selected are: (i) a hard sphere gas (angle and energy independent differential cross
section $\sigma = a^2/4$, where $a$ is the hard sphere radius),
(ii) the Maxwell gas ($\sigma (g,\theta)= m\Gamma(\theta)/2g$ with $m$
being the mass of the heat bath particles, $\Gamma(\theta)$ is an
arbitrary function of $\theta$, and $g$ is the relative velocity),
(iii) chiral pions (for which the $t-$averaged cross section \\ $\sigma =
s/(64\pi^2 f_\pi^4) \left(3 + \cos^2\theta \right)$, where $s$ and $t$ are the usual Mandelstam
variables and $f_\pi $ (93 MeV) is the pion-decay constant, and
(iv) massive pions (for which the differential elastic cross section is taken from
experiments \cite{Bertsch:88} ). Where possible, analytical results are obtained in
either the non-relativistic or extremely relativistic cases.

\section{The Chapman-Enskog Approximation}
\label{chapman}

In this scheme, the local distribution function is expressed in terms of small deviations from equilibrium in terms of hydrodynamic variables and their gradients.
Successive approximations to the transport coefficients are then developed using relativistic kinetic theory.
For elastic scattering of identical particles (obeying Boltzmann statistics), the first approximation to shear viscosity is  given by \cite{LPG:73}
\begin{eqnarray}
 [\eta_s]_1 =\frac{1}{10}\,kT\,\frac{\gamma_0^2}{c_{00}}\,,
\label{shearfirst}\quad {\rm where}\quad \gamma_0 = -10 \hat{h}\,,\,\,\,\hat{h} = \frac{K_3(z)}{K_2(z)}\,,\,\,\,z = \frac{mc^2}{kT}\,\,
{\rm and}\,\,c_{00} = 16\left( w_2^{(2)} - \frac{1}{z}\,w_1^{(2)} + \frac{1}{3z^2}w_0^{(2)} \right)\,. \label{c00massive}
\end{eqnarray}
The quantity $w_i^{(s)}$ is the so-called the relativistic omega integral given by
\begin{eqnarray}
w_{i}^{(s)} &=& \frac{2\pi z^3c}{K_2(z)^2}\int_{0}^{\infty} d\psi\,
\sinh^7  \psi\, \cosh^i\psi\, K_j(2z\cosh\psi)\,\int_{0}^{\pi} d\Theta\, \sin \Theta\, \sigma(\psi,\Theta)\,(1-\cos^s\Theta)~\,,
\label{omega1}
\label{relomega}
\end{eqnarray}
where $j = \frac{5}{3}+\frac{1}{2}\left( -1\right)^i$. The relative and center of mass momenta $g$ and $P$ are given by
\begin{eqnarray}
g = \frac{1}{2}(p_1-p_2)\,,\quad P =
(-p_{\alpha}p^{\alpha})^{1/2}\,,\quad
\sinh \psi = \frac{g}{mc}\,\,,\,\,\textnormal{and} \,\,\,
\cosh \psi = \frac{P}{2mc}~.
\label{omcoeffs}
\end{eqnarray}
The integral involving the differential cross section $\sigma(\psi,\Theta)$ is generally referred to as the transport cross section. The above expressions are readily reduced to their non-relativistic counterparts for
$z\gg 1$ \cite{Kox:76}.

\section{Relaxation Time Approximation}
In this method, the main assumption is that the
effect of collisions is to bring the perturbed distribution
function $ f({\bf x},{\bf p})$ close to the equilibrium distribution function
$f^{eq}({\bf x},{\bf p})$  over a time
$\tau$ which is of order the time required between particle
collisions. The collision integral of the Boltzmann equation can then be written as
%
$ D_c f({\bf x},{\bf p}) = - \frac{f({\bf x},{\bf p}) - f^{eq}({\bf x},{\bf p})}{\tau} \,.$
%
Following closely the formalism described in Refs.~\cite{Gavin:85,Kapusta:09,CK:10},
we restrict our attention to  two-body elastic reactions $a+b \rightarrow c+d$
in a heat bath containing a single species of particles. Employing the notation in Ref. \cite{CK:10}, the shear viscosity is given by \cite{CK:10}
\begin{eqnarray}
 \eta_s &=& \frac{1}{15T}\,\int_0^{\infty}\,\frac{d^3p_a}{(2\pi)^3}\,
\frac{|p_a|^4}{E_a^2}\,\frac{1}{w_a(E_a)}\,f^{eq}_a\,,\,\,\,{\rm where}\,\,\,
f_a^{eq}({\bf x},{\bf p}_a,t) = \frac{1}{{\rm e}^{(E_a-\mu_a)/T} - (-1)^{2s_a}} \,.
\label{etasrelax}
\end{eqnarray}
Above,
$w_a(E_a)$ is the collision frequency which takes the form
\begin{eqnarray}
  w_a(E_a)  &=&  \int \frac{d^3p_b}{(2\pi)^3}\, \,\frac{\sqrt{s(s-4m^2)}}{2E_a\,E_b} \,\frac {1}{2}\sigma_T\,f^{eq}_b\,,
  \label{freqrelax2}
\end{eqnarray}
where $\sigma_T$ is the total cross section. Interactions appear in the
collision frequency through the total cross section.  Here we see the
difference with the Chapman-Enskog approximation which features a
transport cross section that favors right-angled collisions in the
center of mass frame.

\section{Comparison of Results and Conclusion}
\begin{table*}
\caption{The Chapman-Enskog (sec. 2) and relaxation time (sec. 3) shear viscosities of nonrelativistic systems.\label{tab1}}
\begin{tabular}{|l|c|c|c|c|}
\hline
Case & Cross-section & Chapman-Enskog Method & Relaxation Time Method & Chapman-Enskog/Relaxation Time \\
\hline 
Hard-sphere & $\sigma = \frac{a^2}{4}$ & $0.078\,
\sqrt{\frac{m\,k_BT}{\pi}}\,\frac{1}{a^2}$ &
$0.049\,\sqrt{\frac{m\,k_BT}{\pi}}\,\frac{1}{a^2}$ &1.59\\
& & & & \\
Maxwell gas & $\sigma_0 = \frac{m\,\Gamma(\theta)}{2\,g}$ & $\frac{k_BT}{2\,\pi\,\Gamma}$
& $\frac{k_BT}{2\pi\,\Gamma} $ &1.00 \\
\hline
\end{tabular}
\end{table*}

\begin{table*}
\caption{The Chapman-Enskog (sec. 2) and relaxation time (sec. 3) shear viscosities of ultra-relativistic systems..\label{tab2}}
\begin{tabular}{|l|c|c|c|c|}
\hline
Case & Cross-section & Chapman-Enskog Method & Relaxation Time Method & Chapman-Enskog/Relaxation Time \\
\hline 
Hard-sphere & $\sigma_0 = \frac{a^2}{4}$ & $ 1.2\,\frac{k_BT}{\pi\,a^2}\,\frac{1}{c}$ &  $ \frac{8}{5}\,\frac{k_BT}{\pi\,a^2\,c}\,$ & 1.33  \\
& & & & \\
Chiral pions & $\sigma = \frac
{s}{(64\pi^2 f_\pi^4)} $ & $\frac{15\pi}{184}\,\frac{f_{\pi}^4}{T} \frac {1}{\hbar^2c^3}$ & $\frac{12\pi}{25}\,\frac{f_{\pi}^4}{T}\frac {1}{\hbar^2c^3} $ &0.169 \\
& $\times \left(3 + \cos^2 \theta \right)$ & & & \\
\hline
\end{tabular}
\end{table*}

\begin{figure}[t]
\includegraphics[width=9cm]{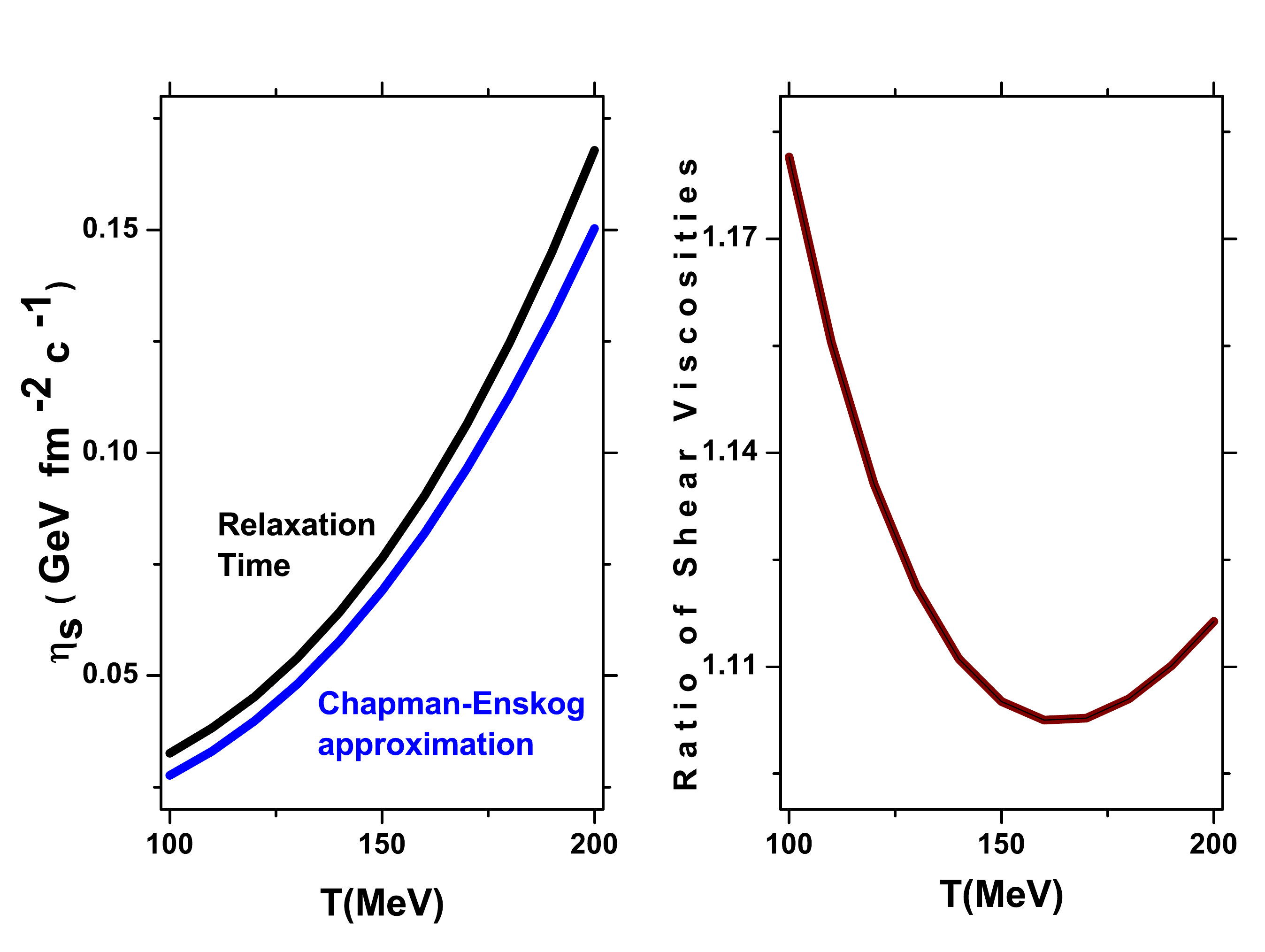}
\caption{Left panel: $\eta_s$ of massive pions from the
relaxation time approximation
and the first order Chapman-Enskog approximation.
Right panel: The ratio of $\eta_s$ from the relaxation time to Chapman-Enskog approximations.
\label{fig1}}
\end{figure}

Table \ref{tab1} shows results for non-relativistic ($z = mc^2/k_BT \gg 1$)
hard sphere  and Maxwell particles.
Results in the ultra-relativistic limit, explored in the cases of the hard sphere
gas \cite{Kox:76} and massless pions \cite{PPVW:93}, are shown in  Table \ref{tab2}.
In the case of massive interacting pions with
experimental cross sections,
 calculations are performed using the
relativistic scheme in Eqs. (\ref{shearfirst}) and (\ref{etasrelax}) as in Refs. \cite{PPVW:93} and \cite{CK:10} (Fig. \ref{fig1}). The results in Tables \ref{tab1} and \ref{tab2}  and those in Fig. \ref{fig1} must be viewed bearing in mind the
difference that exists in the two calculational procedures. The
Chapman-Enskog approximation features the transport cross section with
an angular weight of $(1 - \cos^2 \Theta)$ in first order
calculations. The relaxation time approach lacks this angular
weighting. The angular integral can be performed analytically for the
cases chosen and leads to a factor of 4/3. Even so, it is intriguing
that for the case of Maxwell particles with $\Gamma(\theta)=\Gamma$, the two methods give exactly
the same result. This agreement can be attributed to  the fact that the relative
velocity appearing in the denominator of the cross section is exactly
cancelled by a similar factor occuring in the numerator in both
methods. In the remaining cases, it is clear from the tables that the
energy dependence of the cross sections plays a crucial role in
determining the extent to which results differ between the two
approaches.  This trend persists even with higher order results in the Chapman-Enskog approximation~\cite{Wiranata:2011}.
 In Fig. \ref{fig1}, the first order results of shear viscosity from the
Chapman-Enskog approach are compared with those from the relaxation time
approach (left panel). The right panel shows the ratio of the relaxation time viscosity to that from the Chapman-Enskog viscosity in first order.

The analytical and numerical results of our comparative study reveal that
the extent of agreement (or disagreement) depends sensitively on the
energy dependence of the differential cross sections employed.
Our results (i) call for checks from the more exact Green-Kubo calculations of shear viscosity,
and, (ii) stress the need to combine all available experimental
knowledge concerning
differential cross sections for low mass hadrons and to supplement them
with theoretical guidance for the as yet unknown cross sections so that
the temperature dependent shear viscosity to entropy ratio can be
established for use in viscous hydrodynamics.

\section*{Acknowledgements}
We thank J. I. Kapusta for helpful discussions.
Research support from U. S. DOE grants DE-AC02-05CH11231 and CCNU through
colleges of basic research \& operation of MOE (for A. W.),
from the U. S. DOE grants DE-FG02-93ER-40756 (for A. W. and M. P.) and
DE-FG02-87ER40328 (for P. C.) are gratefully acknowledged.

\bibliographystyle{h-physrev3}
\bibliography{thesisref}

\end{document}